\documentclass[
aps,%
12pt,%
final,%
notitlepage,%
oneside,%
onecolumn,%
nobibnotes,%
nofootinbib,%
superscriptaddress,%
noshowpacs,%
centertags]%
{revtex4}
\usepackage{color}
\begin{document}

\title{Gravitation radiation from the collision of particles\\}

\author{\firstname{M.~A.}~\surname{Misyura}}
\email{max.misyura94@gmail.com}
\affiliation{%
Saint Petersburg State University, Saint Petersburg, Russia
}%


\begin{abstract}
At the present moment, there is an unresolved problem in the physics of cosmic rays. The problem concerns the origin of the ultrahigh energy cosmic rays. One of the possible mechanisms of the particle's birth with such energy is bound to the processes occurring in the active nuclei of the galaxies. In this paper, we have examined a collision of superheavy particles. Using the Weinberg formalism, we have estimated gravitational radiation emitted during the collision in the first approximation. It was shown that, for a certain collision time, the energy lost by the ultrarelativistic particle is less than the value of the energy of the mass center of the collision and the mass of the colliding particles.
\end{abstract}

\maketitle

\section{Introduction}
This study is associated with the fact of the existence of dark matter. The hypothesis expressed in \cite{bib:creation} says that the particles of the dark matter are superheavy, neutral particles with spin equal to zero and with mass of the order of the Grand Unification $10^{15}$ GeV. A possible experimental confirmation of the existence of such heavy dark matter particles is the detection of cosmic rays with ultrahigh energies. The term ultrahigh is associated with their enormous energy, more than $5\times10^{18}$ eV. Today there is a problem. Where can such particles with such high kinetic energy be born? There are various assumptions, for example, it can be intergalactic explosions created in the era of the formation of galaxies, supernovae or gamma bursts. Another assumption is related to supermassive black holes with a mass of the order of millions of solar masses. Such black holes are the nuclei of galaxies. Due to their huge masses, the particles can accelerate up to enormous energies in the black hole accretion disk.  For this reason, there is an opinion that the active galactic nuclei can be the sources of ultrahigh-energy cosmic rays.  Also, there are experimental data which say that the direction of the high-energy detected cosmic rays correlates with the location of supermassive black holes \cite{bib:uhecr}. Moreover, in the works \cite{bib:g1,bib:g2} it was shown that as a result of two-particle scattering, one of the particles can reach ultra-relativistic energies and after that a collision occurs with third particle. The particles can be superheavy particles of the dark matter. The collision can generate protons with sufficient energy, which can reach our planet.  But due to the fact that the relative speed of the particles tends to the speed of light, such particles will emit gravitational waves and thus they will lose energy. Therefore, the aim of this research is the estimation of the energy which is carried away by the gravitational waves at the collision of the particles spinning around the Kerr black hole.

\section{Gravitational Radiation}

We consider particle collisions on a small scale and calculate gravitational radiation in the first approximation. Consequently, we apply the usual expansion of the metric tensor of the flat spacetime
\begin{equation}
g_{ik} = \eta_{ik}+h_{ik}.
\end{equation}
If we represent the energy-momentum tensor as the Fourier integral 
\begin{equation}
\label{2}
T_{ik}(\textbf{x},t) = T_{ik}(\textbf{x},\omega) \exp(-\imath \omega t),
\end{equation} 
it is possible to derive a formula obtained by Wienbegr in \cite{bib:gag}, which characterizes the energy per solid angle $do$ emitted by a system
\begin{equation}
\label{3}
\frac{dE}{do} =  \frac{2G}{c^5} \int\limits^{\infty}_0  \omega^2  \left[T^{ik*}(\textbf{k},\omega) T_{ik}(\textbf{k},\omega) - \frac{1}{2} \left|T^i_i(\textbf{k},\omega)\right|^2\right] d \omega.
\end{equation}
 
Now we examine a system of particles $N$ that initially have a four-momentum $P^i_N$, which collide at $t=0$, and then have a four-momentum $\tilde{P}^i_N $
\begin{equation}
\label{4}
T^{ik}(\textbf{x},t) = \sum_N \frac{P^i_N P^k_N }{P^0_N} \delta^3(\textbf{x}-\textbf{x}_N) \theta(-t)
+\sum_N \frac{\tilde{P}^i_N \tilde{P}^k_N }{\tilde{P}^0_N} \delta^3(\textbf{x}-\tilde{\textbf{x}}_N) \theta(t),
\end{equation}
where $\theta(t)$ is the step function
\begin{equation}
\theta(s) = +1\:s>0, \: \theta(s) =  0,\:s<0,               
\end{equation} 
After the Fourier transform $T^{ik}(\textbf{x},t)$ takes the form 
\begin{equation}
\label{5}
T^{ik}(\textbf{k},\omega) = \frac{c^4}{2 \pi \imath } \sum_N \frac{P^i_N P^k_N \lambda_N}{(P_N \cdot k)}, 
\end{equation}
with $(P_N \cdot k)$ being a scalar product, $\lambda_N$ is the sign factor and $N$ runs over all of the particles in both states
\begin{equation}
\lambda_N = +1 \;  \;\text{in final state},\:  \lambda_N = -1 \;  \;\text{in initial state}.
\end{equation}
We substitute the expression (\ref{5}) in (\ref{3}) and integrate over the directions. Thereby we obtain the formula that characterizes the gravitational radiation per unit frequency interval emitted in a collision
\begin{equation}
\label{7}
\frac{dE}{d \omega} = \frac{G}{2 c  \pi } \sum_{N,M} \lambda_N \lambda_M m_N m_M \frac{1+\beta^2_{NM}}{\beta_{NM}(1-\beta^2_{NM})^{1/2}} \ln\left(\frac{1+\beta_{NM}}{1-\beta_{NM}}\right),
\end{equation}
where $\beta_{NM}$ is the relative speed of the particles $N$ and $M$
\begin{equation}
\beta_{NM} = \left[1-\frac{m^2_N m^2_M}{(P_N \cdot P_M)^2}\right]^{1/2}.
\end{equation}	
Also, this formula in $D$-dimensional flat spacetimes, which was used to calculate gravitational radiation from the collision of a black hole and point particles, can be found in \cite{bib:cardoso}.

We assume that $N=M=2$, $m_N=m_M$  and calculate the maximum limit of radiated energy. The maximum value of the radiated energy corresponds to the case when the relative speed before the collision tends to the speed of light, i.e. the ratio 
\begin{equation}
\label{9}
\frac{m^2_1 m^2_2}{(P_1 \cdot P_2)^2},
\end{equation} 
approaches to zero and $\beta_{12} \rightarrow1$, but after the collision the scalar product $(P_1 \cdot P_2)$ is equal to the product of the masses hence $\beta_{12}=0$. Also we have to cut the $\omega$-integral at $\omega$ of order $1/\Delta t $ since the collision does not occur instantaneously. In this case the equation (\ref{7}) has the following form
\begin{equation}
\label{123}
E =  \frac{G m^2}{ c \Delta t \pi } \frac{1+\beta^2_{12}}{\beta_{12}(1-\beta^2_{12})^{1/2}} \ln\left(\frac{1+\beta_{12}}{1-\beta_{12}}\right).
\end{equation}
Also, if the collision time is equal to the ratio of the Compton length of the particles to the speed of light, the expression (\ref{123}) take the form
\begin{equation}
E =  \left(\frac{m}{M_p}\right)^2 \frac{m c^2}{2 \pi^2} \frac{1+\beta^2_{12}}{\beta_{12}(1-\beta^2_{12})^{1/2}} \ln\left(\frac{1+\beta_{12}}{1-\beta_{12}}\right),
\end{equation}
where $M_p$ is the Planck mass.
  
If the ratio (\ref{9}) is equal to $4.165\times10^{-4}$ which corresponds to the ultrarelativistic case of motion when the energy of the center of mass is close to the grand unification energy $10^{16}$ GeV , $m=10^{15}$ GeV and the collision time is equal to the ratio of the Compton length of superheavy particles to the speed of light, the gravitational radiation from a single collision is:
\begin{equation}
\Delta t = 4.136\times10^{-39} \; \text{s},\:    E =  3.058\times10^{17}  \; \text{eV}.
\end{equation} 
As it can be seen, the energy value is less than the value of the energy of the mass center and the mass of the colliding particles.
It should be noted, that these energy quantities are the maximum possible values that correspond to the case: when the particles before the collision have the relative speed tending to the speed of light but after the collision, this relative speed is zero. However, in the majority of the physical processes, the relative speed both before and after the collision tends to speed of light ($\beta_{12} \rightarrow1$), because the scalar product $(P_1 \cdot P_2)$ is equal roughly  to the energy of the center of mass. Thus, in the expression (\ref{7}), we have the residual of two quantities of equal order. Therefore, the radiated energy should tend to zero, but more accurate conclusions will be made in the next study.  

\section{Conclusion}
As a result of the calculations, it was revealed that the energy lost by the ultrarelativistic particle is eight orders of the magnitude less than the energy of the center of mass. This means that gravitational radiation does not prevent the existence of BSW-resonance (Ba\~{n}ados, Silk and West) in particles collision close to the horizon. Therefore, it can be concluded that cosmic rays with ultrahigh energies can be produced in particle collisions of dark matter with a mass of the order of the Grand Unification scale. This result is an argument in favor of the fact that the active nuclei of galaxies can be one of the possible sources of cosmic rays that exceed the Greisen-Zatsepin-Kuzmin limit.  

\newpage

\begin{acknowledgments}
The author thanks Prof. A. A. Grib for the suggested research subject. Also, the author gratefully acknowledges partial support from the Russian Basic Research Foundation, grant no. 18-02-00461.   
\end{acknowledgments}


\begin{thebibliography}{0}
\expandafter\ifx\csname natexlab\endcsname\relax\def\natexlab#1{#1}\fi
\expandafter\ifx\csname bibnamefont\endcsname\relax
  \def\bibnamefont#1{#1}\fi
\expandafter\ifx\csname bibfnamefont\endcsname\relax
  \def\bibfnamefont#1{#1}\fi
\expandafter\ifx\csname citenamefont\endcsname\relax
  \def\citenamefont#1{#1}\fi
\expandafter\ifx\csname url\endcsname\relax
  \def\url#1{\texttt{#1}}\fi
\expandafter\ifx\csname urlprefix\endcsname\relax\def\urlprefix{URL }\fi
\providecommand{\bibinfo}[2]{#2}
\providecommand{\eprint}[2][]{\url{#2}}

\end{thebibliography}


\section*{References}
\begin{thebibliography}{99}
\bibitem{bib:creation}
A. A. Grib and V. Yu. Dorofeev, {\it Int. J. M. Phys. D} {\bf 3}, 731 (1994). 
\bibitem{bib:uhecr}
HESS Collab. (A. Abramowski {\it et al}.), {\it Nature} {\bf 531}, 476 (2016).
\bibitem{bib:g1}
A. A. Grib and Yu. V. Pavlov, {\it Gravitation and Cosmology} {\bf 17}, 42 (2011).
\bibitem{bib:g2}
A. A. Grib and Yu. V. Pavlov, {\it Astroparticle Physics} {\bf 34}, 581 (2011). 
\bibitem{bib:gag}
S. Weinberg, {\it Gravitation and Cosmology: Principles and Applications of the General Theory of Relativity} (John Wiley and Sons, New York, 1972).

\bibitem{bib:cardoso}
V. Cardoso, \'O. J. C. Dias and J. P. S. Lemos, {\it Phys. Rev. D} {\bf 67}, 064026 (2003).



\end{thebibliography}
\end{document}